\documentclass[preprint,showpacs,preprintnumbers,amsmath,amssymb]{revtex4}

\usepackage{graphicx}
\usepackage{dcolumn}
\usepackage{bm}
\newcommand{\vecvar}[1]{\mbox{\boldmath$#1$}}
\newcommand{\e}{\mbox{e}}

\newcommand{\ssla}{\hspace{-1.5mm}/\hspace{-0.5mm}/} 
\newcommand{\dsla}{\hspace{-0.5mm}/\hspace{-0.5mm}/} 

\begin{document}

\preprint{PRE/001}

\title{Novel time-saving first-principles calculation method for electron-transport properties}

\author{Yoshiyuki Egami}
\affiliation{Nagasaki university Advanced Computing Center, Nagasaki University, Nagasaki 852-8521, Japan}

\author{Kikuji Hirose}
\affiliation{Division of Precision Science \& Technology and Applied Physics, Graduate School of Engineering, Osaka University, Suita, Osaka 565-0871, Japan}

\author{Tomoya Ono}
\affiliation{Division of Precision Science \& Technology and Applied Physics, Graduate School of Engineering, Osaka University, Suita, Osaka 565-0871, Japan}

\date{\today}

\begin{abstract}
We present a time-saving simulator within the framework of the density functional theory to calculate the transport properties of electrons through nanostructures suspended between semi-infinite electrodes. By introducing the Fourier transform and preconditioning conjugate-gradient algorithms into the simulator, a highly efficient performance can be achieved in determining scattering wave functions and electron-transport properties of nanostructures suspended between semi-infinite jellium electrodes.
To demonstrate the performance of the present algorithms, we study the conductance of metallic nanowires and the origin of the oscillatory behavior in the conductance of an Ir nanowire. It is confirmed that the $s$-$d_{z^2}$ channel of the Ir nanowire exhibits the transmission oscillation with a period of two-atom length, which is also dominant in the experimentally obtained conductance trace.
\end{abstract}

\pacs{
}
\maketitle

\section{\label{sec:level1}Introduction}
The understanding and control of the transport properties of electrons through nanostructures are important subjects for the development of new electronic devices. In this decade, numerous experimental and theoretical investigations have been performed on the atomic geometries and transport properties of nanostructures~\cite{agrait}. In the theoretical studies, the nonequilibrium Green's function method within the tight-binding approach~\cite{keldysh,taylor,vega} has been widely employed to describe the quantum transport in nanoscale devices. While this method is efficient for investigating the transport properties of large systems, the completeness of a basis set is always a concern and the precise description of tunneling transport is hard work.

To obtain exact theoretical knowledge on electron-transport properties, Fujimoto and Hirose presented the overbridging boundary-matching (OBM) method~\cite{fujimoto,icp}, formulated by the real-space finite-difference (RSFD) approach~\cite{fd,rsfd1,rsfd5,rsfd8,ono1,icp} within the framework of the density functional theory~\cite{kohn,dft}. Since the system is divided into the equally spaced grid points in the RSFD approach and the wave functions and potentials are directly defined on the grid points, one can strictly treat systems with arbitrary boundary conditions and eliminate difficulties arising from the incompleteness of a basis set. Therefore, the OBM method enables us to examine tunneling transport~\cite{tunnel} as well as ballistic transport~\cite{fujimoto,icp,OBMworksC,OBMworksJ,smit,onoJPC,okamoto,sim,okano,furuya,tawara,tsukamoto,ohnishi,lang1,nkobayashi,egamiPRB,egami_jim} in nanostructures suspended between semi-infinite electrodes with a high degree of accuracy. However, in the OBM method, a large computational cost is required to calculate the Green's function necessary for determining the scattering wave functions. Recently, to avoid this computational hard work, Kong, Tiago, and Chelikowsky made an improvement to the OBM method~\cite{Kong}, enabling the scattering wave functions to be obtained by solving a set of simultaneous equations without the need to calculate the Green's function. We call this method the improved OBM (IOBM) method. However, the IOBM method is still inconvenient when electrodes have a large cross-sectional area in the transition region or are made of multivalent materials.

In this paper, we propose efficient algorithms to solve the simultaneous equations arising in the IOBM procedure and demonstrate electron-transport calculations for nanoscale junctions. To exemplify the advantages of our algorithms, we apply them to investigate the electron-transport properties of a single-row Na nanowire and find that the algorithms can give reasonable numerical solutions within a short CPU time.
We also examine the electron-transport properties of single-row Ir and Au monoatomic nanowires and explore why the even-odd oscillation is dominant in the experimentally obtained conductance traces. The results indicate that the even-odd oscillation is insensitive to the structural deformation of the nanowire, while oscillations with a longer period than two-atom length are easily affected by structural deformation. This implies that only the even-odd oscillation can survive and the other oscillation patterns are cancelled out in experiments.

The remainder of this paper is organized as follows: Section~\ref{sec:level2} gives the details of the computational scheme used to develop an efficient IOBM simulator to determine the electron-transport properties of nanostructures. In Sec.~\ref{sec:level3}, the performance of our method is demonstrated. Furthermore, in Sec.~\ref{sec:level4}, we adopt the method to examine the transport properties of Ir and Au atomic nanowire models and discuss the origin of the oscillatory behavior of conductance. Finally, our works are summarized in Sec.~\ref{sec:level5} and mathematical details are described in Appendix.

\section{\label{sec:level2}Computational Formalism}
We treat a system including a nanostructure suspended between semi-infinite electrodes as shown in Fig.~\ref{fig1}. Here, the $x$ ($y$) and $z$ coordinates are chosen to be parallel and perpendicular to the electrode surface, respectively, and the case of incident electrons propagating from the left electrode is considered. The system is infinite in the $z$ direction and periodic in the $x$ and $y$ directions.
The transition region is divided by grid points with equal spacing of $h_\mu=L_\mu / N_\mu$ in the conventional OBM and IOBM methods based on the RSFD approach~\cite{icp}, where $L_\mu$ and $N_\mu$ are the length and the number of grid points in the $\mu$ direction ($\mu=x$, $y$, and $z$) of the transition region, respectively. The second-order derivative of the wave function $\psi$ in the Kohn--Sham equation is described as
\begin{eqnarray}
\label{eqn01}
\frac{\partial^2}{\partial \vecvar{r}^2} \psi(\vecvar{r}) &\!\!\approx\!\!& \sum_{n=-N_f}^{N_f} \!\! \Bigl\{ c_{x,n} \psi(x_i+nh_x,y_j,z_k) + c_{y,n} \psi(x_i,y_j+nh_y,z_k) + c_{z,n} \psi(x_i,y_j,z_k+nh_z) \Bigr\}, \nonumber \\
\end{eqnarray}
where $N_f$ represents the parameter determining the order of the finite-difference approximation and the coefficients $c_{\mu,n}$ are described in Ref.~7. 
The Kohn--Sham equation is written in a discretized matrix form as
\begin{eqnarray}
-B_z^{\dagger} \Psi(z_{k-1}) + \left[E-H(z_k;\vecvar{\kappa}_{\dsla})\right]\Psi(z_{k}) -B_z \Psi(z_{k+1})= 0 &, & \nonumber \\
(k=-\infty,...,-1,0,1,...,\infty) & &
\label{eqn02}
\end{eqnarray}
where $B_z= -\frac{1}{2 h_z^2} I$ with $I$ being the $N_{xy}$-dimensional unit matrix ($N_{xy}=N_x \times N_y$), $H(z_k;\vecvar{\kappa}_{\dsla})$ denotes the $N_{xy}$-dimensional block-tridiagonal matrix including the potential on the $x$--$y$ plane at $z=z_k$, and $\vecvar{\kappa}_{\dsla}=(\kappa_x,\kappa_y)$ is the lateral Bloch wave vector within the first Brillouin zone. Here, $\Psi(z_k)$ is the set of values of the wave function on the $x$--$y$ plane at $z=z_k$, $\{\psi(x_i,y_j,z_k):i=1,\cdots,N_x,~j=1,\cdots,N_y\}$.
For simplicity, the central finite-difference formula ($N_f=1$) is adopted, and the extension to the case of a higher-order finite-difference representation is straightforward.
In the transition region ($z_0 \leq z \leq z_{N_z+1}$), we rewrite Eq.~(\ref{eqn02}) in a matrix representation as
\begin{equation}
\left[E-\hat{H}_T\right]\left[
\begin{array}{c}
\Psi(z_{0})   \\
\Psi(z_{1})   \\
\vdots        \\
\Psi(z_{N_z})   \\
\Psi(z_{N_z+1}) \\
\end{array}
\right]
=
\left[
\begin{array}{c}
B_z^{\dagger} \Psi(z_{-1})   \\ 
    0         \\
\vdots        \\
    0         \\
B_z \Psi(z_{N_z+2}) \\
\end{array}
\right], \label{eqn03}
\end{equation}
where $E$ is the Kohn--Sham energy and $\hat{H}_T$ is the Hamiltonian of the truncated part of the system sandwiched between the planes at $z=z_{0}$ and $z=z_{N_z+1}$. $\hat{H}_T$ forms the $N_x \times N_y \times (N_z+2)$-dimensional block-tridiagonal matrix given by
\begin{equation}
\hat{H}_T= \left[
\begin{array}{ccccc}
H(z_{0};\vecvar{\kappa}_{\dsla}) & B_z & & & 0 \\
B_z^{\dagger} & H(z_{1};\vecvar{\kappa}_{\dsla}) & B_z & & \\
 & \ddots & \ddots & \ddots & \\
 & & B_z^{\dagger} & H(z_{N_z};\vecvar{\kappa}_{\dsla}) & B_z \\
0 & & & B_z^{\dagger} & H(z_{N_z+1};\vecvar{\kappa}_{\dsla})
\end{array}
\right].
\label{eqn04}
\end{equation}
Here and hereafter, the dependence of the Hamiltonian $\hat{H}_T$ on the lateral Bloch wave vector $\vecvar{\kappa}_{\dsla}$ is omitted from the suffix of variables to avoid complication.

In the original OBM procedure~\cite{fujimoto,icp}, the scattering wave functions are evaluated by employing the Green's function $\hat{\mathcal{G}}_T(E)$ which is defined as the inverse matrix of $[E-\hat{H}_T]$.
When the $(i,j)$ block-matrix element of $\hat{\mathcal{G}}_T(E)$ is denoted as $\mathcal{G}_{i,j}$, Eq.~(\ref{eqn03}) is rewritten as 
\begin{eqnarray}
\left[
\begin{array}{c}
\Psi(z_{0})   \\
\Psi(z_{1})   \\
\vdots        \\
\Psi(z_{N_z+1}) \\
\end{array}
\right] &=& \left[
\begin{array}{rcl}
B_z^{\dagger}\mathcal{G}_{  0,0}\Psi(z_{-1}) & + & B_z\mathcal{G}_{  0,N_z+1}\Psi(z_{N_z+2}) \\
B_z^{\dagger}\mathcal{G}_{  1,0}\Psi(z_{-1}) & + & B_z\mathcal{G}_{  1,N_z+1}\Psi(z_{N_z+2}) \\
 & \vdots &  \\
B_z^{\dagger}\mathcal{G}_{N_z+1,0}\Psi(z_{-1}) & + & B_z\mathcal{G}_{N_z+1,N_z+1}\Psi(z_{N_z+2}) \\
\end{array}
\right].
\label{eqn05}
\end{eqnarray}
Note that it is only necessary to work out the 0th and $(N_z+1)$th block-column elements of $\hat{\mathcal{G}}_T(E)$, $\mathcal{G}_{i,0}$, and $\mathcal{G}_{i,N_z+1}$ ($i=0,1,\cdots,N_z+1$). Since their calculation requires a computational load proportional to $O(N_{xy}^2 \times N_z)$ with an iterative method such as the conjugate gradient (CG) method, a large computational load is required to adopt the OBM method to large systems.

On the other hand, in the IOBM method~\cite{Kong}, the scattering wave functions of electrons propagating from the left electrode are evaluated by solving the following simultaneous equations for each incident wave function $\Phi^{in}_L(z_k)$ with an iterative method:
\begin{equation}
\left[E-\hat{H}_T-\tilde{H}\right]\left[
\begin{array}{c}
\Psi(z_{0})   \\
\Psi(z_{1})   \\
\vdots        \\
\Psi(z_{N_z+1}) \\
\end{array}
\right]
=
\left[
\begin{array}{c}
B_z^{\dagger}\Phi^{in}_L(z_{-1})-\Sigma^r_L(z_0)\Phi^{in}_L(z_0) \\ 
    0         \\
\vdots        \\
    0         \\
\end{array}
\right].
\label{eqn07}
\end{equation}
Equation~(\ref{eqn07}) is obtained from Eq.~(\ref{eqn03}) using the scattering boundary condition, i.e.,
\begin{eqnarray}
\Psi(z_k)=\left\{
\begin{array}{l}
\displaystyle{\Phi^{in}_L(z_k) + \sum_{i=1}^{N_{xy}} r_i \Phi^{ref}_i(z_k)} \hspace{4mm} (k \le 0) \\
\displaystyle{\sum_{i=1}^{N_{xy}} t_i \Phi^{tra}_i(z_k)} \hspace{4mm} (k \ge N_z+1) \hspace{8mm}
\end{array}
\right.,
\label{eqn06}
\end{eqnarray}
where $r_i$ ($t_i$) is the reflection (transmission) coefficient, and $\Phi^{ref}_i(z_k)$ $\left(\Phi^{tra}_i(z_k)\right)$ is the generalized Bloch state in semi-infinite electrodes for the reflected (transmitted) electrons. Here,
\begin{equation}
\tilde{H}
=
\left[
\begin{array}{ccccc}
\Sigma^r_L(z_0) & 0      & \cdots & \cdots & 0                   \\ 
 0              & 0      & \ddots &        & \vdots              \\
\vdots          & \ddots & \ddots & \ddots & \vdots              \\
\vdots          &        & \ddots & 0      & 0                   \\
 0              & \cdots & \cdots & 0      & \Sigma^r_R(z_{N_z+1}) \\
\end{array}
\right].
\label{eqn07-2}
\end{equation}
Note that $\Sigma^r_{L}$ and $\Sigma^r_{R}$ are the retarded self-energy matrices in the left and right electrodes, respectively, and are evaluated by
\begin{eqnarray}
\Sigma^r_L(z_0)&=&B_z^{\dagger}Q^{ref}(z_{-1})[Q^{ref}(z_0)]^{-1},
\label{eqn17-1}
\end{eqnarray}
and
\begin{eqnarray}
\Sigma^r_R(z_{N_z+1})&=&B_zQ^{tra}(z_{N_z+2})[Q^{tra}(z_{N_z+1})]^{-1},
\label{eqn17-2}
\end{eqnarray}
where $Q^A(z_k)$ ($A=$~{\it ref} and {\it tra}) are $N_{xy}$-dimensional matrices consisting of $\{\Phi_i^A(z_k)\}$, i.e., $Q^A(z_k)=[\Phi_1^A(z_k), \cdots , \Phi_{N_{xy}}^A(z_k)]$ (see Sec.~9.3 in Ref.~6). 
The generalized Bloch states inside the semi-infinite electrodes are constituted by propagating Bloch waves with real wave vectors and evanescent waves with complex ones.
Because the evanescent waves behave as exponential functions and decay during their transmission from deep inside the left electrode, only right-propagating Bloch waves must be taken into account as incident waves $\Phi^{in}_L$.
The scattering wave functions for electrons propagating from the right electrode are also described in a similar manner.

Since the retarded self-energy matrix is an $N_{xy}$-dimensional matrix, the maximum order of the computational cost of solving Eq.~(\ref{eqn07}) is $O(N_{in} \times N_{xy}^2)$ owing to the multiplications of $\Sigma^r_{L}(z_{0}) \times \Psi(z_{0})$ and $\Sigma^r_{R}(z_{N_z+1}) \times \Psi(z_{N_z+1})$, where $N_{in}$ is the number of incident waves.
Although $[E-\hat{H}_T-\tilde{H}]$ is a non-Hermitian matrix, we can save a reasonable amount of computational time compared with that required for the original OBM scheme when $N_{in}$ is much smaller than $N_z$. On the other hand, as $N_{in}$ increases, this procedure for calculating the scattering wave functions consumes a larger CPU time than the original OBM scheme, because the convergence of the CG method for a non-Hermitian matrix is slow.
Therefore, this procedure is not suitable when the system includes electrodes having a large cross-sectional area or consisting of multivalent materials, since $N_{in}$ is proportional to the lengths of sides of the supercell and the number of valence electrons within the electrodes.

We now present a novel procedure for efficiently solving Eq.~(\ref{eqn07}) in the case of jellium electrodes.
The jellium-electrode approximation has been successfully applied to the interpretation of electron-transport properties with less computational load~\cite{tsukamoto,okamoto,onoJPC,furuya,lang1,nkobayashi,OBMworksJ}.
The self-energy matrices of jellium electrodes are independent of $z_k$, and those in the left and right electrodes are the same, i.e., $\Sigma^r_L(z_k)=\Sigma^r_R(z_k)\equiv\Sigma^r$.
Thus, the component of the retarded self-energy matrix $\Sigma^r$ for the grid point $(\vecvar{r}_{\ssla,\ell},\vecvar{r}_{\ssla,\ell^\prime})$ on the $x$--$y$ plane in the case of $N_f=1$ (the central finite-difference case) is analytically given by
\begin{eqnarray}
\label{eqn08}
\Sigma^r_{\ell,\ell^\prime}&=&-\frac{1}{2h_z^2N_{xy}}\sum_{\vecvar{G}_{\dsla,\nu}} \exp\left[i(\vecvar{G}_{\dsla,\nu}+\vecvar{\kappa}_{\dsla})\cdot(\vecvar{r}_{\dsla,\ell}-\vecvar{r}_{\dsla,\ell^\prime})\right] \cdot \exp(ik_{z,\nu}h_z),
\end{eqnarray}
where $\vecvar{G}_{\dsla,\nu}=(G_{\nu_x},G_{\nu_y})$ represents the lateral reciprocal lattice vector $\displaystyle \left(\frac{2\pi}{L_x}\nu_x,\frac{2\pi}{L_y}\nu_y\right)$ with $\nu_x$ and $\nu_y$ being integers, and $k_{z,\nu}$ is the $z$ component of the wave vector in the RSFD scheme defined as
\begin{eqnarray}
k_{z,\nu} &=& \left\{
\begin{array}{lcl}
\displaystyle{\frac{\cos^{-1}(-\chi_\nu)}{h_z}} & \mbox{for propagating waves} & (|\chi_\nu|<1) \vspace{5mm} \\
\displaystyle{\frac{i\cosh^{-1}(-\chi_\nu)}{h_z}} & \mbox{for evanescent waves} & (|\chi_\nu|>1)
\end{array}
\right.
\label{eqn08.5}
\end{eqnarray}
with
\begin{eqnarray}
\chi_{\nu}&=&h_z^2E-\frac{h_z^2}{h_x^2}\bigl(1-\cos(G_{\nu_x}+\kappa_x)h_x\bigr)-\frac{h_z^2}{h_y^2}\bigl(1-\cos(G_{\nu_y}+\kappa_y)h_y\bigr)-1.
\label{eqn09}
\end{eqnarray}
In Eq.~(\ref{eqn08.5}), the branch of the function $\cos^{-1}(z)$ ($\cosh^{-1}(z)$) is so chosen that $k_{z,\nu}$ ($-ik_{z,\nu}$) is a positive value.
The derivation of Eqs.~(\ref{eqn08})--(\ref{eqn09}) is given in Appendix.

The product of the matrix $\Sigma^r$ and the vector $\Psi(z_{k})$, $\displaystyle{\sum_{\ell^\prime} \Sigma^r_{\ell,\ell^\prime}\psi(\vecvar{r}_{\ssla,\ell^\prime},z_k)}$, is in a convolution form of the 2-dimensional Fourier transform, and it can be most easily calculated in the momentum space. Each matrix element of the Fourier transformed retarded self-energy matrix, $\tilde{\Sigma}^r_{\nu,\nu^\prime}$, is given by
\begin{eqnarray}
\tilde{\Sigma}^r_{\nu,\nu^\prime} &=& \frac{1}{N_{xy}} \sum_{\vecvar{r}_{\ssla,\ell}} \sum_{\vecvar{r}_{\ssla,\ell^\prime}} \exp\left[-i(\vecvar{G}_{\dsla,\nu}+\vecvar{\kappa}_{\dsla})\cdot\vecvar{r}_{\ssla,\ell}\right] \exp\left[i(\vecvar{G}_{\dsla,\nu^\prime}+\vecvar{\kappa}_{\dsla})\cdot\vecvar{r}_{\ssla,\ell^\prime}\right] \Sigma^r_{\ell,\ell^\prime}.
\label{eqn18-1}
\end{eqnarray}
By applying the orthogonality of the plane waves, one obtains the following expression of the diagonalized self-energy matrix:
\begin{eqnarray}
\tilde{\Sigma}^r_{\nu,\nu^\prime} &=& -\frac{1}{2h_z^2} \delta_{\nu\nu^\prime} \exp(ik_{z,\nu}h_z).
\label{eqn18-2}
\end{eqnarray}
The columnar vector $\tilde{\Psi}(z_k)$ consists of $N_{xy}$ values of the Fourier transformed wave function of $\Psi(z_k)$ as
\begin{eqnarray}
\tilde{\psi}_{\nu}(z_{k}) &=& \frac{1}{N_{xy}}\sum_{\vecvar{r}_{\ssla,\ell}}\exp\left[-i(\vecvar{G}_{\dsla,\nu}+\vecvar{\kappa}_{\dsla})\cdot\vecvar{r}_{\ssla,\ell}\right] \psi(\vecvar{r}_{\ssla,\ell},z_{k}).
\label{eqn19}
\end{eqnarray}
Because the number of grid points $\vecvar{r}_{\ssla,\ell}$ corresponds to that of the reciprocal lattice vectors $\vecvar{G}_{\dsla,\nu}$ and the off-diagonal elements of the Fourier transformed retarded self-energy matrix $\tilde{\Sigma}^r$ are zero according to Eq.~(\ref{eqn18-2}), calculating the product of $\tilde{\Sigma}^r$ and $\tilde{\Psi}(z_k)$ requires a computational load of $O(N_{in} \times N_{xy})$.
The Fourier transform of $\Psi(z_k)$ and the inverse Fourier transform of $\tilde{\Sigma}^r \times \tilde{\Psi}(z_k)$ require $O(N_{in} \times N_{xy}\log N_{xy})$ operations when the fast Fourier transform (FFT) algorithm is employed. Thus, the overall scaling of the critical part of the calculations is improved from $O(N_{in} \times N_{xy}^2)$ to $O(N_{in} \times N_{xy}\log N_{xy})$.

Next, we introduce a preconditioning CG (PCG) method into the solver to accelerate convergence since the total CPU time is proportional to the number of iterations.
If the preconditioner $\hat{P}$ is similar to $\bigl[E-\hat{H}_T-\tilde{H}\bigr]^{-1}$, the spectral property of the matrix $\hat{P} \times \bigl[E-\hat{H}_T-\tilde{H}\bigr]$ can contribute to rapid convergence.
The incomplete Cholesky preconditioners are effective at accelerating convergence and are commonly employed, but they require more computations per iteration and larger amounts of memory in general because the preconditioners are not sparse.
One might consider taking an easily calculated and/or easily stored matrix $\hat{P}$ as an approximation to $\bigl[E-\hat{H}_T-\tilde{H}\bigr]^{-1}$.
Although the Jacobi preconditioner, in which the preconditioner $\hat{P}$ is chosen to be the diagonal matrix in terms of the reciprocal of the diagonal elements of $\bigl[E-\hat{H}_T-\tilde{H}\bigr]$, is one of the simplest and most useful forms, more sophisticated preconditioning is required for faster convergence.
Here, we propose a method for improving $\hat{P}$ using the Green's function of the Laplacian operator.
We postulate that the Laplacian of the kinetic operator in $\bigl[E-\hat{H}_T-\tilde{H}\bigr]$ is dominant and use the inverse matrix of the discretized Laplacian in the RSFD approach.
Since the Green's function of the Laplacian operator is represented by $\displaystyle{\frac{1}{{|\vecvar{r}_i-\vecvar{r}_j|}}}$, the inverse matrix of the discretized Laplacian is approximated as $\displaystyle{\frac{1}{|\vecvar{r}_i-\vecvar{r}_j|}}$, where $\vecvar{r}_{i(j)}$ is the position of the $i(j)$th grid point. To avoid the numerical difficulties due to this matrix not being sparse and diverging at $\vecvar{r}_i=\vecvar{r}_j$, we employ the following truncated matrix as the preconditioner:
\begin{eqnarray}
\hat{P}\gamma(\vecvar{r}) &=& C_0\gamma(\vecvar{r})+\sum_{\vecvar{r}^{\prime}=(\pm h_x,\pm h_y,\pm h_z)} C_{1} \gamma(\vecvar{r}+\vecvar{r}^{\prime}),
\label{eqn10}
\end{eqnarray}
where $\gamma$ is the residual vector of the CG iteration, and $C_0$ and $C_1$ are coefficients used in the preconditioning, which are discussed later. We confirmed that this preconditioning can successfully reduce the CG iteration count required to solve the Poisson equation~\cite{icp}.

\section{\label{sec:level3}Performance Test}
\begin{figure}[p]
\begin{center}
\includegraphics[width=8.0cm]{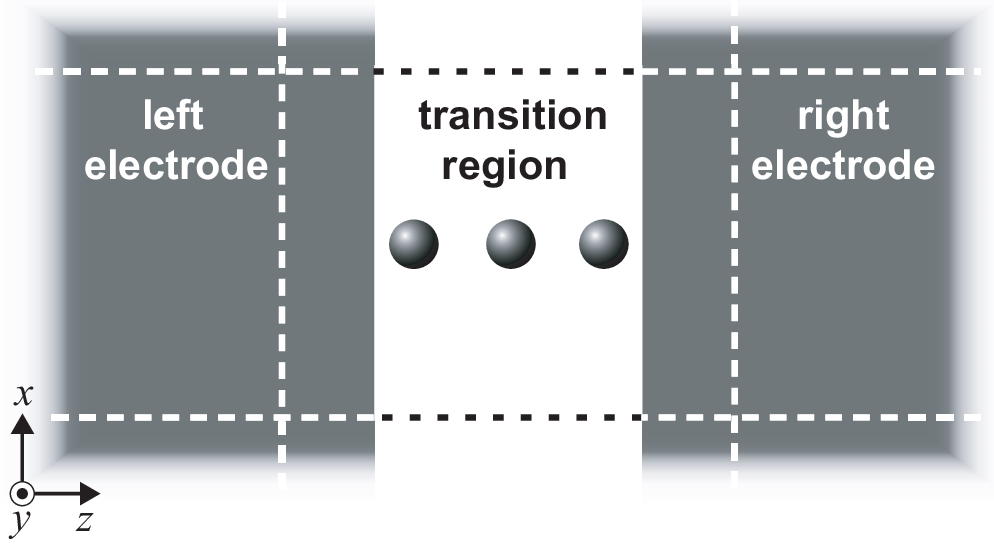}
\caption{ 
Schematic view of ballistic transport system. The transition region is suspended between the left and right semi-infinite electrodes. In the $x$ and $y$ directions, periodic boundary conditions are imposed. The dashed lines represent the boundaries of the supercell employed to determine the Kohn--Sham effective potential and the transition region employed to calculate the transport properties. The shaded area represents jellium electrodes.
}
\label{fig1}
\end{center}
\end{figure}

In order to demonstrate the performance of the IOBM method incorporated with the FFT and PCG algorithms, the electron-transport properties of single-row Na nanowire models suspended between semi-infinite electrodes are examined (see Fig.~\ref{fig1}). The number of atoms consisting the nanowire, $N_{atom}$, is varied between 3 and 8, and the nanowires are directly attached to structureless jellium electrodes. The interatomic distance is $d$~$(=\sqrt{3}a_0/2)$, where $a_0$~(=8.11~a.u.) is the lattice constant of Na bulk. The distance between the edge atom of the nanowire and the surface of the jellium electrode is $\sqrt{2}a_0/4$, so as to correspond to a [110] Na strand. The Wigner-Seitz radius of the jellium electrodes is taken to be $r_s=3.99$~a.u. The exchange-correlation effects are treated by the local density approximation~\cite{lda} and the interaction between electrons and atomic cores is described by the norm-conserving pseudopotentials of Troullier and Martins~\cite{tmpp,norm}.
To determine the Kohn--Sham effective potential, a conventional supercell is employed under a periodic boundary condition in all directions, which is represented by a rectangle denoted by dashed lines in Fig.~\ref{fig1}; the lengths of the supercell are $L_{x(y)}=25.66$~a.u. in the $x(y)$ direction and $L_z=(N_{atom}-1) \! \cdot \! d$+42.12~a.u. in the $z$ direction.
The numbers of grid points in the $x(y)$ and $z$ directions are set to be $N_{x(y)}=36$ and $N_z=60+10 \times (N_{atom}-1)$, respectively, and the number of incident waves, $N_{in}$, is 13 at the Fermi level. In this case, the ratio of the maximum order for the calculation of $\Sigma^r \times \Psi(z_{0(N_z+1)})$ with the FFT algorithm to that without adopting the Fourier transform is $\frac{N_{in} \times N_{xy}\log N_{xy}}{N_{in} \times N_{xy}^2}=0.0024$. The conductance of the nanowire system at the limits of zero temperature and zero bias is determined by the Landauer-B\"uttiker formula~\cite{buttiker}.
The numerical examination is carried out on a workstation with a 3.25~GHz Intel\textregistered~Xeon\textregistered~processor using four different solvers: (a) the original OBM method, (b) the IOBM method, (c) the IOBM method with the FFT algorithm, and (d) the IOBM method with the FFT and PCG algorithms.
In solver (d), we adopt $C_0=1$ and $C_1=\exp(-\alpha)$ with $\alpha$ being a positive real number.

\begin{figure}[p]
\begin{center}
\includegraphics[width=8.0cm]{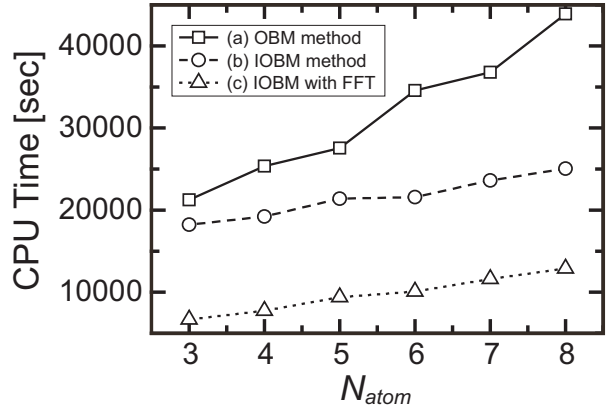}
\caption{ 
CPU time required to calculate the conductances of single-row Na nanowires as a function of the number of atoms constituting the nanowires, $N_{atom}$. (a) original OBM method, (b) IOBM method, and (c) IOBM method with the FFT algorithm.
}
\label{fig:Time1}
\end{center}
\end{figure}

\begin{figure}[p]
\begin{center}
\includegraphics[width=8.0cm]{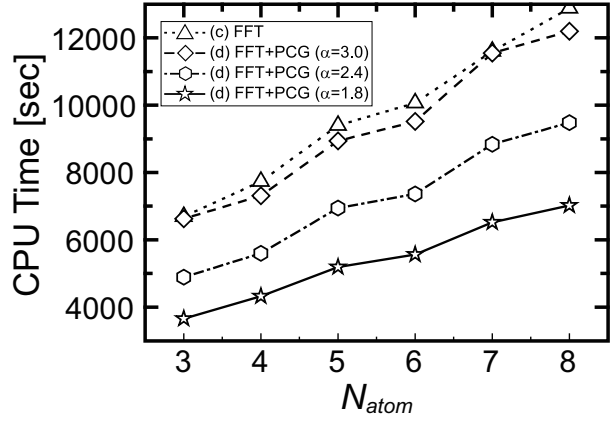}
\caption{ 
CPU time required to calculate the conductances of single-row Na nanowires as a function of the number of atoms constituting the nanowires, $N_{atom}$, employing the IOBM method with the FFT and PCG algorithms. (c) and (d) correspond to the solvers described in the text.
}
\label{fig:Time2}
\end{center}
\end{figure}

CPU time versus $N_{atom}$ for solvers~(a), (b), and (c) is displayed in Fig.~\ref{fig:Time1}. Note that solver~(c) results in faster convergence than the others.
To investigate the effect of the parameter of the PCG algorithm, Fig.~\ref{fig:Time2} shows CPU time versus $N_{atom}$ for solver~(d) with $\alpha=3.0$, $2.4$, and $1.8$. The CPU time of solver~(c) is also plotted in Fig.~\ref{fig:Time2} for comparison.
Table~\ref{tab2} shows the average number of iterations required to obtain convergent solutions for incident waves. The average numbers of iterations for solver~(d) with $\alpha=3.0$, $2.4$, and $1.8$ are approximately 1.3, 1.7, and 2.3 times smaller than that for solver~(c), respectively.
The combination of the FFT and PCG algorithms enables us to significantly reduce the CPU time required for the calculation of the electron-transport properties of nanostructures with moderate memory consumption.

\begin{figure}[p]
\begin{center}
\includegraphics[width=8.0cm]{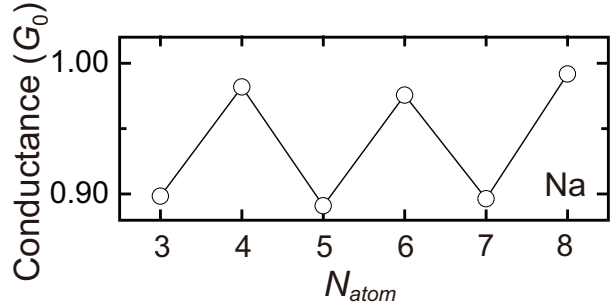}
\end{center}
\caption{Conductance of single-row Na nanowires as a function of the number of atoms constituting the nanowires, $N_{atom}$.}
\label{fig:conductanceNa}
\end{figure}

Figure~\ref{fig:conductanceNa} shows the conductance of Na nanowires in a unit of ${\rm G_0}$ (${\rm G_0}=2e^2/h$, where $e$ is the electron charge and $h$ is Planck's constant). The conductance is $\sim 1~{\rm G_0}$ and exhibits oscillatory behavior with a period of two-atom length, i.e., the well-known even-odd oscillation.
In previous theoretical studies, it was demonstrated that the conductance of a Na nanowire is not significantly different from 1~G$_0$ and that it oscillates with respect to the nanowire length~\cite{lang1,nkobayashi,sim,tsukamoto,egami_jim,egamiPRB}. Our result is in good agreement with those of previous studies.

\section{\label{sec:level4}Applications}

In 2003, Smit {\it et al.}~\cite{smit} found, using mechanically controllable break junctions, that the conductance of Ir and Pt monoatomic nanowires manifests oscillatory behavior with a period of two-atom length, similarly to Na and Au ones. de la Vega {\it et al.}~\cite{vega} carried out a tight-binding calculation on the electron-transport properties of Ir, Pt, and Au nanowires and reported that additional oscillation patterns with a longer period and larger amplitude than those obtained in Ref.~2 can be observed in the conductance traces of Ir and Pt nanowires. Although it is intuitively expected that patterns with a large amplitude are dominant in the conductance traces, no experimental evidence of such oscillation patterns has been measured up to now. Recently, one of the present authors (T.~O.) has examined the transport properties of a Pt nanowire by first-principles calculation and claimed that the even-odd oscillation is due to the low sensitivity of the transmission oscillation of the $s$-$d_{z^2}$ channel to the spatial deformation of the nanowire~\cite{onoJPC}. On the other hand, in the case of the Ir nanowire, the transmission of the $s$-$d_{z^2}$ channel oscillates with a longer period than that of the Pt nanowire according to a tight-binding calculation~\cite{vega}. It is of interest to explore using first-principles calculations whether the even-odd oscillation of the conductance trace of the Ir nanowire is caused by the $s$-$d_{z^2}$ channel. In addition, since 5$d$ transition metals have the multiple valence electrons, $N_{in}$ for Ir electrodes is larger than that for Na ones, which causes a significant increase in computational cost. For this reason, we apply the method described in the previous section to examine the transport properties of the Ir nanowire.

\begin{figure}[p]
\begin{center}
\includegraphics[width=8.0cm]{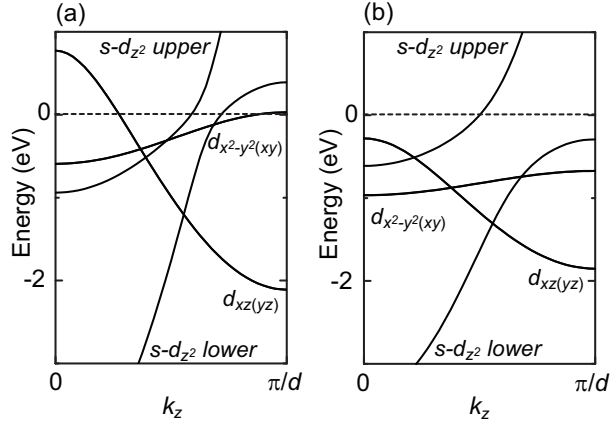}
\end{center}
\caption{Energy band structures of infinite (a) Ir and (b) Au wires. The zero of energy is chosen to be the Fermi level.}
\label{fig:inf-band}
\end{figure}

We first calculate the electronic structures of infinite straight Ir and Au wires of equal interatomic distance $d$~$(=a_f/\sqrt{2})$, where $a_f$ is the lattice constant of fcc bulk ($a_f=7.25$ and $7.71$~a.u. for Ir and Au, respectively). The grid spacing $h$ is set to be $\sqrt{3}a_f$/34 and a denser grid spacing of $h$/3 is employed in the vicinity of nuclei by the augmentation of the double-grid technique~\cite{icp,ono1}. The supercell contains an atom under a periodic boundary condition, and the size of the supercell is $L_x=L_y=46h$ and $L_z=d$, where $L_x(L_y)$ and $L_z$ are the lengths of the supercell in the $x(y)$ directions perpendicular to the wire and in the $z$ direction parallel to the wire, respectively. The exchange-correlation effects are treated by the local density approximation~\cite{lda} and the interaction between electrons and atomic cores is described by the norm-conserving pseudopotentials of Troullier and Martins~\cite{tmpp,norm}. The integration over the Brillouin zone along the wire-axis direction is performed by the sampling of 80 equidistant {\it k}-points. We verified that the increase in the numbers of grid points and $k$-points did not affect our conclusion. Figure~\ref{fig:inf-band} shows the energy band structures of the infinite Ir and Au wires. The bands are labeled according to the atomic orbitals mainly constituting the bands. Only the upper $s$-$d_{z^2}$ band crosses the Fermi level in the case of the Au wire, whereas the other $d$ bands also cross the Fermi level in the case of the Ir wire. These results are consistent with experimental results, in which the maximum conductance of the Au nanowire is 1~${\rm G_0}$ while the conductance of the Ir nanowire exceeds 1~${\rm G_0}$~\cite{smit}. In addition, it is well known that the oscillatory behavior of the conductance is led by the quantum-mechanical wave character of the electrons, and that the period of the oscillation is given by $\pi/k_z$ at the intersection between the band and the Fermi level, where $k_z$ is the component of the wave vector along the wire axis in the infinite atomic wire~\cite{egamiPRB}. Since $k_z$ for the upper $s$-$d_{z^2}$ channel is $\pi/2d$ for both the Au wire and the Ir one, the transmission of these channels is expected to exhibit even-odd oscillation.

We next examine the electron-transport properties of Ir and Au nanowires suspended between semi-infinite electrodes to ensure that the even-odd oscillation is observed in the transmission of the upper $s$-$d_{z^2}$ channel. Hereafter, the superscripts $Ir$ and $Au$ denote the parameters for Ir and Au nanowires, respectively. The interatomic distances are the same as those of the infinite wires. The distance between the edge atom of the nanowire and the surface of the jellium electrode is 2$a_f/\sqrt{6}$. The Wigner-Seitz radii of the jellium electrodes are taken to be $r_s^{Ir}=1.36$~a.u. and $r_s^{Au}=1.35$~a.u. so that the elements of the electrodes correspond to those of the nanowires. The grid spacing of $d/16$ is employed and $N_{atom}$ is varied between 3 and 8.
The lengths of the transition region in the $x$ and $y$ directions are $L_x^{Ir}=L_y^{Ir}=15.07$~a.u. and $L_x^{Au}=L_y^{Au}=15.38$~a.u., and that in the $z$ direction is $L_z=(N_{atom}+4) \! \cdot \! d$~a.u. The numbers of grid points in the $x(y)$ direction are set to be $N_{x(y)}^{Ir}=46$ and $N_{x(y)}^{Au}=48$, and that in the $z$ direction is $N_z=16 \times (N_{atom}+4)$. The number of incident waves, $N_{in}$, is 37 for both models. In this case, the ratios of the maximum order for the calculation of $\Sigma^r \times \Psi(z_{0(N_z+1)})$ with the FFT algorithm to that without the Fourier transform are $\frac{N_{in} \times N_{xy}^{Ir}\log (N_{xy}^{Ir})}{N_{in} \times (N_{xy}^{Ir})^2}=0.0016$ and $\frac{N_{in} \times N_{xy}^{Au}\log (N_{xy}^{Au})}{N_{in} \times (N_{xy}^{Au})^2}=0.0015$ for the Ir and Au nanowires, respectively. The other computational conditions are the same as those for the infinite wires.

\begin{figure}[p]
\begin{center}
\includegraphics[width=8.0cm]{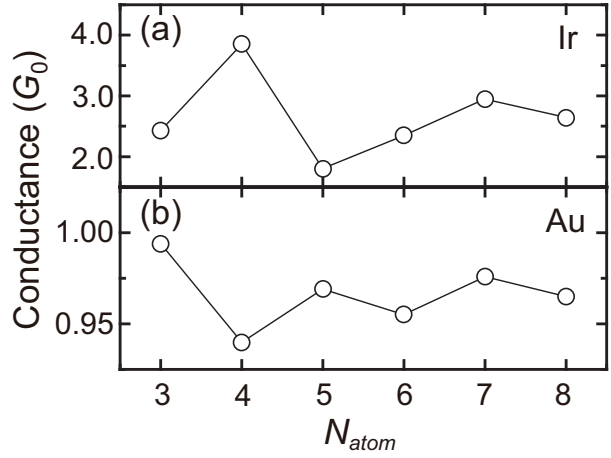}
\end{center}
\caption{Conductances of (a) Ir and (b) Au nanowires as a function of the number of atoms constituting the nanowires, $N_{atom}$.}
\label{fig:conductanceIrAu}
\end{figure}

Figure~\ref{fig:conductanceIrAu} shows the conductance of the Ir and Au nanowires in the unit of ${\rm G_0}$. The conductance trace of the Au nanowire exhibits even-odd oscillation depending on $N_{atom}$ with an amplitude of $\sim 0.02$~${\rm G_0}$, while no oscillatory behavior is observed in the conductance trace of the Ir nanowires. To obtain a deeper understanding of the oscillatory behavior, the evolution of the channel decomposition of the conductance trace is shown in Table \ref{tbl:tbllir}, in which the quantum numbers of the eigenchannels correspond to those in the case of infinite wires. The eigenchannels are computed by diagonalizing the Hermitian matrix {\bf T}$^\dag${\bf T}~\cite{nkobayashi}, where {\bf T} is the transmission matrix. Even-odd oscillation clearly emerges in the transmission of the upper $s$-$d_{z^2}$ channel of the Ir nanowire, and the other $d$ channels exhibit oscillation with a longer period and a larger amplitude of $\sim 0.5~{\rm G_0}$. According to a previous theoretical study on Pt nanowires~\cite{onoJPC}, the contributions of the other $d$ channels become negligible upon being averaged over thousands of scans in experiments. Thus, we can conclude that the even-odd oscillation in the conductance trace of the Ir nanowire observed in the experiment~\cite{smit} is attributed to the transmission oscillation of the upper $s$-$d_{z^2}$ channel. 

\section{\label{sec:level5}Conclusion}
We have presented algorithms for efficiently solving a set of simultaneous equations described in the IOBM method proposed by Kong, Tiago, and Chelikowsky~\cite{Kong}. In the case of structureless jellium electrodes, the self-energy matrices are analytically given and are diagonal in momentum space. These spectral properties of the self-energy matrices enable us to reduce the computational cost required for the multiplication of the self-energy matrices and wave functions. Furthermore, a solver that successfully accelerated the CG algorithm has been developed by introducing a preconditioning technique. In this way, we developed an extremely efficient simulator. To demonstrate the performance of the presented algorithms, we applied them to calculate the transport properties of an Ir nanowire attached to semi-infinite electrodes. The upper $s$-$d_{z^2}$ channel of the Ir nanowire mainly contributes to the electron transport and gives rise to the oscillation of the transmission with a period of two-atom length.
Our accelerated algorithms are expected to lead to a greater understanding of the physics underlying electron transport through nanostructures using large computational models.
\section*{\label{sec6}Acknowledgments}
This research was partially supported by the Nagasaki University Tenure Track Program and a Grant-in-Aid for the Global COE in Osaka University ``Center of Excellence for Atomically Controlled Fabrication Technology'' from the Ministry of Education, Culture, Sports, Science and Technology.
The numerical calculation was carried out with the computer facilities at the Institute for Solid State Physics at the University of Tokyo, the Information Synergy Center at Tohoku University, and the Cybermedia Center at Osaka University.


\newpage

\begin{table}[p]
\caption{Average CG iteration counts required to calculate scattering wave functions for incident waves in single-row sodium nanowires employing the IOBM method with the FFT and PCG algorithms.}
\begin{tabular}{c|c|c|c|c}
\hline
\hline
\hspace{1mm}$N_{atom}$\hspace{1mm} & {\footnotesize without PCG} & \ \ $\alpha=3.0$ \ \ & \ \ $\alpha=2.4$ \ \ & \ \ $\alpha=1.8$ \ \ \\
\hline
3 & 13936.46 & 10864.38 & 8241.92 & 6107.30 \\
4 & 13852.77 & 10738.76 & 8184.76 & 6030.84 \\
5 & 14688.00 & 11518.92 & 8918.15 & 6620.00 \\
6 & 14058.38 & 10995.69 & 8336.00 & 6326.38 \\
7 & 14639.15 & 12036.07 & 9111.92 & 6750.23 \\
8 & 14818.23 & 11597.76 & 8794.30 & 6613.76 \\
\hline
\hline
\end{tabular}
\label{tab2}
\end{table}

\begin{table}[p]
\begin{center}
\caption{Channel transmissions of Ir nanowires.}
\label{tbl:tbllir}
\begin{tabular}{c|cccc} \hline\hline
\hspace{3mm}$n$\hspace{3mm} & upper $s$-$d_{z^2}$ \hspace{1mm} & \hspace{1mm} lower $s$-$d_{z^2}$ & \hspace{3mm} $d_{xz(yz)}$ \hspace{3mm} & \hspace{3mm} $d_{x^2-y^2(xy)}$ \hspace{3mm} \\ \hline
3 & 0.854 & 0.016 & 0.763 & 0.016 \\
4 & 0.973 & 0.011 & 0.935 & 0.497 \\
5 & 0.727 & 0.018 & 0.510 & 0.019 \\
6 & 1.000 & 0.020 & 0.656 & 0.010 \\
7 & 0.838 & 0.056 & 0.979 & 0.048 \\
8 & 0.975 & 0.116 & 0.569 & 0.206 \\ \hline\hline
\end{tabular}
\\
\end{center}
\end{table}

\appendix*\section{}
\subsection*{Retarded self-energy matrix of a semi-infinite jellium electrode}

As shown in Eqs.~(\ref{eqn17-1}) and (\ref{eqn17-2}), the retarded self-energy matrices $\Sigma^r_{L}$ and $\Sigma^r_{R}$ are obtained by using the generalized Bloch states $Q^{ref}(z_k)$ and $Q^{tra}(z_k)$.
On the other hand, the self-energy matrices are also derived from the retarded Green's function $\mathcal{G}^{r}_{\{L,R\}}$ in the semi-infinite electrodes (see p.160 of Ref.~6) as
\begin{eqnarray}
\Sigma^r_L(z_0)&=&B_z^{\dagger}\mathcal{G}^{r}_L(z_{-1},z_{-1};E,\vecvar{\kappa}_{\dsla})B_z,
\label{eqnA-01}
\end{eqnarray}
and
\begin{eqnarray}
\Sigma^r_R(z_{N_z+1})&=&B_z\mathcal{G}^{r}_R(z_{N_z+2},z_{N_z+2};E,\vecvar{\kappa}_{\dsla})B_z^{\dagger}.
\label{eqnA-02}
\end{eqnarray}

\begin{figure}[p]
\begin{center}
\includegraphics[width=8.0cm]{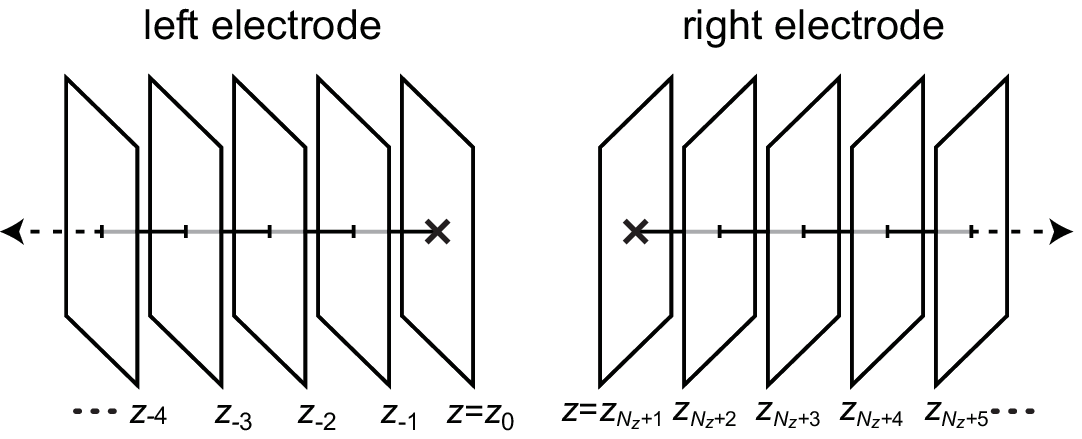}
\caption{ 
Schematic view of discretized 3-dimensional semi-infinite systems.
}
\label{figA-1}
\end{center}
\end{figure}

In this appendix, we introduce the analytical derivation of the retarded Green's function in the left-hand semi-infinite jellium electrode. That in the right-hand one is derived in the same manner. For simplicity, the lateral Bloch vector $\vecvar{\kappa}_{\dsla}$ is taken to be zero.
We deal with the model of a free electron in which the discretized space is semi-infinite in the $z$ direction and periodic in the $x$ and $y$ directions~(see Fig.~\ref{figA-1}). The grid points are denoted by $\vecvar{r}_l=(x_{l_x},y_{l_y},z_{l_z})=(l_x h_x,l_y h_y, l_z h_z)$, where $l_{\mu}=-N_{\mu}/2, ..., -1,0,1, ..., N_{\mu}/2-1$, and $h_{\mu}$ and $N_{\mu}$ are the grid spacing and the total number of the grid points in the $\mu$ direction ($\mu=x,y$), respectively. We first assume a finite system in the $z$ direction, i.e., $l_z=-N_z-1,-N_z,\cdots,-1,0$ and impose the zero boundary condition on the wave function, $\phi_m(z_{-N_z-1})=\phi_m(z_{0})=0$, and then a semi-infinite system will be represented by taking the limit of $N_z \rightarrow \infty$.
Adopting the central finite-difference formula for the second-order derivative, we write the Schr\"odinger equation for a free electron according to Eq.~(\ref{eqn01}) as
\begin{eqnarray}
-\frac{1}{2}\sum_{n=-1}^{1} \Bigl\{ c_{x,n} \phi_m(x_{l_x}+nh_x,y_{l_y},z_{l_z}) + c_{y,n} \phi_m(x_{l_x},y_{l_y}+nh_y,z_{l_z}) \Bigr. && \nonumber \\
\Bigl. + c_{z,n} \phi_m(x_{l_x},y_{l_y},z_{l_z}+nh_z) \Bigr\} &=& E_m \phi_m(x_{l_x},y_{l_y},z_{l_z}) \nonumber \\
\label{eqnA-03}
\end{eqnarray}
with $c_{\mu,0}=-2/h_\mu^2$ and $c_{\mu,\pm 1}=1/h_\mu^2$ $(\mu=x,y,z)$.
In a finite system, the Hamiltonian has discrete eigenvalues and therefore the Green's function
\begin{eqnarray}
\mathcal{G}(\vecvar{r}_l,\vecvar{r}_{l^\prime};Z)&=&\sum_{m}\frac{\phi_m(\vecvar{r}_l)\phi^{*}_m(\vecvar{r}_{l^\prime})}{Z-E_m}
\label{eqnA-04}
\end{eqnarray}
exhibits only simple poles at the positions of the eigenvalues in the complex $Z$ plane, where $ \{ E_m \}$ is the set of the eigenvalues and $\{ \phi_m(\vecvar{r}) \}$ is the complete orthonormal set of the eigenfunctions of the Hamiltonian.
In this model, the eigenfunctions and eigenvalues are analytically described as
\begin{eqnarray}
\phi_m(x_{l_x},y_{l_y},z_{l_z}) &=& \sqrt{\frac{2}{N_{xy}(N_z+1)}}\exp\Bigl\{i(G_{\nu_x}x_{l_x}+G_{\nu_y}y_{l_y})\Bigr\}\sin{\beta_{\nu_z}z_{l_z}} \\
\mbox{and} \ \ \ E_m &=&\frac{1}{h^2_x}(1-\cos{G_{\nu_x}h_x})+\frac{1}{h^2_y}(1-\cos{G_{\nu_y}h_y})+\frac{1}{h^2_z}(1-\cos{\beta_{\nu_z}h_z}),
\end{eqnarray}
respectively, where $\displaystyle{\beta_{\nu_z}=\frac{\pi}{(N_z+1)h_z}\nu_z}$ and $N_{xy}=N_x \times N_y$. Thus, the Green's function of this finite system is given by
\begin{eqnarray}
\mathcal{G}_L(\vecvar{r}_l,\vecvar{r}_{l^\prime};Z)&=&\frac{2}{N_{xy}(N_z+1)}\sum^{\frac{N_x}{2}-1}_{\nu_x=-\frac{N_x}{2}} \sum^{\frac{N_y}{2}-1}_{\nu_y=-\frac{N_y}{2}} {\exp\bigl\{iG_{\nu_x}(x_{l_x}-x_{l_x^\prime})\bigr\}\exp\bigl\{iG_{\nu_y}(y_{l_y}-y_{l_y^\prime})\bigr\}} \nonumber \\
&& \times \sum^{N_z}_{\nu_z=1} \frac{\sin{\beta_{\nu_z}z_{l_z}}\sin{\beta_{\nu_z}z_{l_z^\prime}}}{Z-E_\nu}.
\label{eqnA-05}
\end{eqnarray}

We go on to the derivation of the Green's function for a free electron in the semi-infinite discretized space. By carrying out a limiting procedure $N_z \rightarrow \infty$ while keeping $h_z$ constant in (\ref{eqnA-05}), we have
\begin{eqnarray}
\mathcal{G}_L(\vecvar{r}_l,\vecvar{r}_{l^\prime};Z)&=&\frac{h_z^2}{2\pi N_{xy}}\sum^{\frac{N_x}{2}-1}_{\nu_x=-\frac{N_x}{2}} \sum^{\frac{N_y}{2}-1}_{\nu_y=-\frac{N_y}{2}} {\exp\bigl\{iG_{\nu_x}(x_{l_x}-x_{l_x^\prime})\bigr\}\exp\bigl\{iG_{\nu_y}(y_{l_y}-y_{l_y^\prime})\bigr\}} \nonumber \\
&&\times \int^{\pi}_{-\pi} \frac{\exp\{i\theta(l_z-l_z^\prime)\}-\exp\{i\theta(l_z+l_z^\prime)\}}{h^2_z Z-\bigl\{ \frac{h_z^2}{h_x^2}(1-\cos{G_{\nu_x}h_x})+\frac{h_z^2}{h_y^2}(1-\cos{G_{\nu_y}h_y})+(1-\cos\theta) \bigr\} } d \theta. \nonumber \\
\label{eqnA-06}
\end{eqnarray}
To evaluate the integral in Eq.~(\ref{eqnA-06}), we transform it to an integral over the complex variable $\omega$ ($=\e^{i \theta}$) along the unit circle as
\begin{eqnarray}
\mathcal{G}_L(\vecvar{r}_l,\vecvar{r}_{l^\prime};Z) &=& \frac{h_z^2}{i\pi N_{xy}}\sum^{\frac{N_x}{2}-1}_{\nu_x=-\frac{N_x}{2}} \sum^{\frac{N_y}{2}-1}_{\nu_y=-\frac{N_y}{2}} {\exp\bigl\{iG_{\nu_x}(x_{l_x}-x_{l_x^\prime})\bigr\}\exp\bigl\{iG_{\nu_y}(y_{l_y}-y_{l_y^\prime})\bigr\}} \nonumber \\
&&\times \oint_{|\omega|=1} \frac{\omega^{|l_z-l_z^\prime|}-\omega^{|l_z+l_z^\prime|}}{\bigl(\omega-\omega_1(Z)\bigr)\bigl(\omega-\omega_2(Z)\bigr)}d\omega,
\label{eqnA-07}
\end{eqnarray}
where
\begin{eqnarray}
\omega_{1}(Z)&=&-\chi_\nu(Z) + \sqrt{\chi_\nu(Z)^2 -1 } \nonumber \\
\omega_{2}(Z)&=&-\chi_\nu(Z) - \sqrt{\chi_\nu(Z)^2 -1 } \nonumber \\
\chi_\nu(Z)&=&h_z^2Z-\frac{h_z^2}{h_x^2}(1-\cos{G_{\nu_x}h_x})-\frac{h_z^2}{h_y^2}(1-\cos{G_{\nu_y}h_y})-1.
\label{eqnA-08}
\end{eqnarray}
It follows that $\omega_1\omega_2=1$. Only when the pole\index{pole} $\omega=\omega_1$ or $\omega_2$ in (\ref{eqnA-07}) exists inside the unit circle in the $\omega$~plane, does it contribute to the integral.
In a similar manner as shown in p.149-151 of Ref.~6, the retarded Green's function is derived as
\begin{eqnarray}
\mathcal{G}^r_L(\vecvar{r}_l,\vecvar{r}_{l^\prime};E) &=& \frac{h_z^2}{i N_{xy}}\sum^{\frac{N_x}{2}-1}_{\nu_x=-\frac{N_x}{2}} \sum^{\frac{N_y}{2}-1}_{\nu_y=-\frac{N_y}{2}} {\exp\bigl\{iG_{\nu_x}(x_{l_x}-x_{l_x^\prime})\bigr\}\exp\bigl\{iG_{\nu_y}(y_{l_y}-y_{l_y^\prime})\bigr\}} \nonumber \\
&&\times \frac{1}{\sin{k_{z,\nu}h_z}}\Bigl[ \exp\{ik_{z,\nu}|z_{l_z}-z_{l_z^\prime}|\}-\exp\{ ik_{z,\nu}|z_{l_z}+z_{l_z^\prime}| \} \Bigr]
\label{eqnA-09}
\end{eqnarray}
with $k_{z,\nu}$ defined by Eq.~(\ref{eqn08.5}).
Finally, the retarded Green's function at the surface of the left-hand semi-infinite electrode is represented by
\begin{eqnarray}
\mathcal{G}^r_L(x_{l_x},x_{l_x^\prime},y_{l_y},y_{l_y^\prime},z_{-1},z_{-1},;E) &=& -\frac{2h_z^2}{N_{xy}}\sum^{\frac{N_x}{2}-1}_{\nu_x=-\frac{N_x}{2}} \sum^{\frac{N_y}{2}-1}_{\nu_y=-\frac{N_y}{2}} {\exp\bigl\{iG_{\nu_x}(x_{l_x}-x_{l_x^\prime})\bigr\}} \nonumber \\
&&\times \exp\bigl\{iG_{\nu_y}(y_{l_y}-y_{l_y^\prime})\bigr\} \exp( ik_{z,\nu}h_z ).
\label{eqnA-10}
\end{eqnarray}
The above-mentioned derivation of the Green's function is straightforwardly extended to the case of a higher-order finite-difference approximation.

\end{document}